\documentclass[amssymb,prl,twocolumn,showpacs]{revtex4}
\usepackage{amsmath,amsthm,amsfonts}
\usepackage{epsf}
\usepackage{amssymb}
\usepackage{graphicx}

\newcommand{\be}{\begin{eqnarray}                       }
\newcommand{\ee}{\end{eqnarray}       }
\newcommand{\ket}[1]{\left |#1\right \rangle}
\newcommand{\bra}[1]{\left \langle #1\right|}
\newcommand{\eps}{\varepsilon}

\begin{document}
\title{Vibrational Coherences in Nano-Elastic Tunneling}
\author{Hannes H\"ubener $^1$ and Tobias Brandes $^2$}

\affiliation{1-1. Institut f\"ur Theoretische Physik, Universit\"at Hamburg, D-20355 Hamburg, Germany }
\affiliation{2-Institut f\"ur Theoretische Physik, Technische Universit\"at Berlin, D-10623 Berlin, Germany. 
}
\begin{abstract}
Charging a nano-scale oscillator by single electron tunneling leads to an effective double-well potential due to image charges. We combine exact numerical diagonalizations with generalized Master equations and show that the resulting quantum tunneling of the mechanical degree of freedom can be visualized in the electronic current noise spectrum.

\end{abstract}
\date{\today{ }}
\pacs{
72.70.+m 
73.23.Hk 
73.63.Kv 
85.35.Gv	
85.85.+j	
}
\maketitle

Nanoelectromechanical systems (NEMS) display a variety of interesting dynamical effects due to the coupling between mechanical and electronic degrees of freedom. Some recent examples are oscillators with linear coupling to single \cite{koch} or multiple \cite{rodriguesPRL} electronic levels, or systems with non-linear coupling such as quantum shuttles \cite{shuttling}.

A common feature of NEMS is the  possibility to study quantum-{\em mechanics} in the true meaning of the word, i.e. quantum coherent behavior of objects (molecules or mechanical resonators \cite{exp_resonators}) that are big on the scale of smaller units (electrons, atoms) \cite{blencowe}.
In this Letter, we show that coherent quantum tunneling occurs in a small vibrating resonator that is driven by  stochastic image charges and a corresponding switching between  harmonic and non-harmonic (Coulomb-like) potentials generated by  single electron transport. For certain parameters,  the resonator tunnels back and forth between two spatially separated positions which can be visualized by  measuring the noise spectrum of the electrons.

\begin{figure}[ht]\centering
		\includegraphics[width=7.5cm]{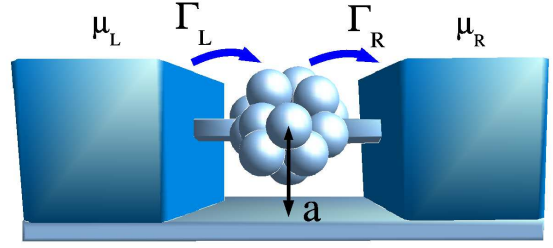}
\caption{Quantum dot in a nanoscale resonator close to a surface: single electrons (de)charging a nanoscale resonator  at rates $\Gamma_{L} (\Gamma_{R})$ lead to a switching between a harmonic and a non-linear  potential due to the Coulomb attraction of the charged resonator towards a surface image charge.}
\label{figMod}
 \end{figure}

{\em Model .-}
Our model resonator oscillates in $x$-direction with frequency $\omega_0$ and mean square displacement $l_0$ in a  harmonic oscillator potential that is modified by tunneling of a single electron through a nearby quantum dot or molecular level. Coulomb attraction of the resonator towards an image charge induced at a distance $2|\mathbf{a}|$ in a nearby surface occurs by charging with  single electrons (cf. Fig.\ref{figMod}), e.g.,  in a molecular junction, in quantum dots embedded into a vibrating beam \cite{weigPRL}, or in larger quantum dots where charging with an additional electron attracts the  electron `droplet'  towards a nearby metallic gate.
We thus assume two  electronic `dot' states $|0\rangle$ and $|1\rangle$ (with zero or one transport electron),  and an additional Coulomb potential on the resonator in the 
occupied state,  leading to 
the effective Hamiltonian of the dot-resonator
\be \label{hamilton}
H_{\rm dot}\equiv \frac{\mathbf{p}^2}{2M} + \frac{1}{2}M\omega_0^2(\mathbf{a}-\mathbf{r})^2 - d^{\dagger}d^{\phantom{\dagger}}\frac{e^2}{4 \pi \eps_0 \eps_r 2\mathbf{r\cdot\hat{u}}},
\ee
where $M\equiv \hbar/l_0^2\omega_0$ is the effective oscillator mass of the resonator, $\mathbf{\hat{u}}$ the unit vector from the dot to the image charge, and $d^\dagger$ the creation operator of an electron on the dot, where for simplicity we disregard the electron's spin degree of freedom here and in the following.

We start our calculation by an exact diagonalization of $H_{\rm dot}$. In the three-dimensional problem, the resonator motion in $x$-direction decouples from the motion perpendicular to the $x$-axis for both charged and uncharged states, thus leaving us with an effectively one-dimensional problem: the uncharged vibrational states $|n,0\rangle$ simply are harmonic oscillator eigenstates with respect to the equilibrium position $a$. Non-harmonic corrections to the uncharged resonator potential in Eq. (\ref{hamilton}) can easily be incorporated but do not lead to drastic modifications of our results.

On the other hand, the charged states $|n,1\rangle$ are obtained by expansion into the eigenbasis of $l=0$ hydrogen atom eigenfunctions (weighted Laguerre polynomials) $\psi_n(x)=\sqrt{\frac{(n-1)!}{2n(n!)}}\left(\frac{2\epsilon}{n}\right)^{\frac{3}{2}}e^{-\frac{\epsilon}{n}x}x L^1_{n-1}\left(\frac{2\epsilon}{n}x\right)$ and subsequent numerical diagonalization. Here, $x$ is measured from the surface in units of $a$. 

\begin{figure}[t]\centering
		\includegraphics[width=\columnwidth]{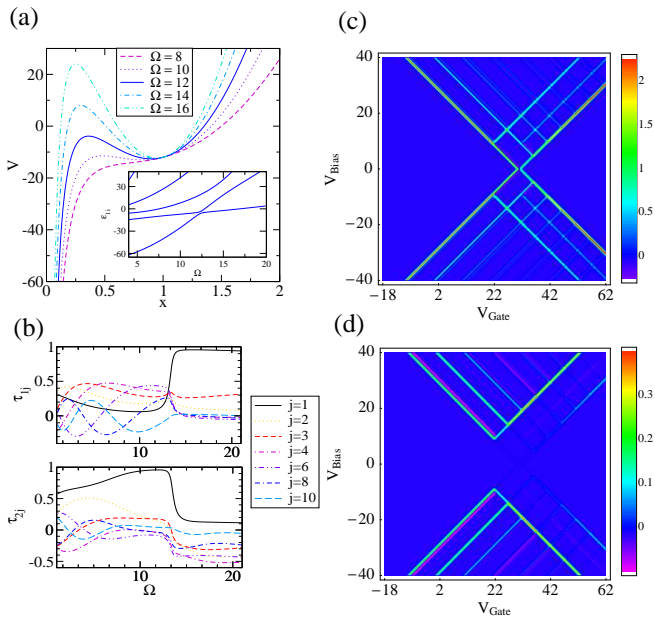}
		\caption{(color online). \textbf{(a)} Potential of an oscillator with image charge effect depending on the oscillator frequency $\Omega$ and the eigenvalue spectrum of the correspondong Hamiltonian (for $\epsilon = 12$) (inset). Level anti-crossing occurs at $\Omega = 13.4$. \textbf{(b)}  Franck-Condon overlaps of the eigenfunctions of the charged and uncharged oscillator as a function of $\Omega$. There are two different regimes before and after the anti-crossing point. \textbf{(c)} $dI/dV$ as a function of gate and bias voltage for $\epsilon = 12$ and $\Omega = 11$ without the effect of the Franck-Condon factors ($\tau_{ij}=1$) showing the usual conductance diamonds. \textbf{(d)} As (c) but with Franck-Condon factors: Current is surpressed for low bias and cannot be restored by changing the gate (Franck-Condon blockade). Negative differential conductance occurs as violet lines in the transport spectrum (c.f color scale)}\label{figPotSpecFcDiam}
 \end{figure}

The model has two dimensionless parameters, 
\begin{eqnarray}\label{parameters}
\Omega \equiv \frac{M\omega_0}{\hbar}a^2 = \frac{a^2}{l_0^2},\quad \epsilon\equiv\frac{e^2 M a}{8\hbar^2\pi\varepsilon_0 \eps_r}= \Omega \frac{E_a}{\hbar\omega_0},
\end{eqnarray}
where $\Omega$ determines the distance from the wall relative to the harmonic oscillator (zero-point fluctuation) length scale $l_0$, and $\epsilon$ is given by $\Omega$ times the typical Coulomb (charge - image charge) energy,
\begin{eqnarray}\label{parameterEa}
E_a \equiv \frac{e^2}{4\pi\varepsilon_0 \eps_r 2a},
\end{eqnarray}
relative to the harmonic oscillator phonon energy $\hbar \omega_0$. Both parameters determine the shape of $V(x)$  
(cf. Fig. \ref{figPotSpecFcDiam}(a))  and the resulting vibrational eigenstates $\ket{n,\nu}$, where $\nu=1(\nu=0)$ denotes (un)charged resonator states and $n$ is the vibrational quantum number. In the following, we  measure all energies (rates) in units of $(\hbar)\omega_0/\Omega$.

An important feature of the corresponding spectrum of the energy eigenvalues  $\eps_{\nu n}$ for $\nu=1$ is the appearance of an avoided level crossing  (Fig. \ref{figPotSpecFcDiam}(a) inset near $\Omega=13.4$). This can be traced back to the effective double-well structure of $V(x)$ for coupling strengths $\epsilon > 9$. In this regime, the charged oscillator can be approximated by a two-level system at low energies, and one thus expects  quantum coherent behavior due to tunneling between two states approximately localized around $x=a$ and $x=0$ (i.e. near the surface). 
 In the following, we demonstrate that this vibrational tunneling can be made visible by electronic transport through the resonator, using Wigner function representations of the stationary density matrix and the frequency-dependent electronic current noise spectrum $S(\omega)$. 

{\em Method .-}
We use the usual tunneling Hamiltonian $H_T=\sum_{k\alpha}V_{k,\alpha} c_k^{\dagger}d + H. c. $ with left and right ($\alpha=L,R$)  electronic reservoirs at chemical potentials $\mu_L$ and $\mu_R$ described by the free electron Hamiltonian $H_{\rm res}=\sum_{k\alpha} \varepsilon_k c^{\dagger}_{k\alpha} c^{\phantom{\dagger}}_{k\alpha}$. 
We describe quantum transport by formulating a generalized Master equation for the reduced dot-resonator density operator $\rho_t$ that is suitable for strong Coulomb blockade (two electronic states) and weak reservoir coupling. This approach describes only first order tunneling processes, i.e. neglecting co-tunneling effects. Importantly and in contrast to the usual rate equation method \cite{koch,mitraPRB}, we fully retain all vibrational coherences in the resulting Master equation, which in the usual Born-Markov approximation is obtained by taking vibrational matrix elements of the Liouvillian \cite{brandesReview}, leading to  
\begin{widetext}
\be\label{master1}\nonumber
\bra{n,1}\dot{\rho}_t\ket{m,1} =-i(\eps_{1n}-\eps_{1m})\rho_{n_1 m_1} - \frac{1}{2}\sum_\alpha\{\!& -
\sum_{j,j'}\gamma_\alpha(\eps_{1n}-\eps_{0j})\tau_{nj} \tau_{mj'}
\rho_{j_0 j'_0} + \sum_{i,j}\overline{\gamma}_\alpha(\eps_{1i}-\eps_{0j})\tau_{nj}
\tau_{ij} \rho_{i_1 m_1} -\\
  &-\sum_{j,j'}\gamma_\alpha(\eps_{1m}-\eps_{0j'})\tau_{nj} \tau_{mj'}
\rho_{j_0 j'_0}+ \sum_{i,j} \overline{\gamma}_\alpha(\eps_{1i}-\eps_{0j})\tau_{ij}
\tau_{m j}\rho_{n_1 i_1}\}
\ee
\end{widetext}
and a corresponding equation for the uncharged states. The indices $j$ and $j'$ run over the uncharged oscillator states and $i$ over the charged states. Note that no coherence between charged and uncharged states occurs as must be. Here, we defined the rates $\gamma_\alpha(\eps) = \Gamma_\alpha f_\alpha(\eps)$  for tunneling into the dot, the rates $\overline{\gamma}_\alpha(\eps)=\Gamma_\alpha(1- f_\alpha(\eps))$ for tunneling out of the dot where the coupling strength to the leads is defined as $\Gamma_\alpha=\sum_{k}V_{k,\alpha}^2 2\pi\delta(\eps-\eps_k)$. Also, we use a short hand notation for the matrix elements $\rho_{n_\nu m_\mu}\equiv\bra{n,\nu}\rho_t\ket{m,\mu}$ and $f_\alpha(\eps)$ for the Fermi function at $\mu_\alpha+\eps$. 

The Franck-Condon factors $\tau_{ij}$ naturally occur by sandwiching $\rho_t$ with the vibrational states. They are defined as the overlap integral of the vibrational wavefunctions $\langle x \ket{n,\nu}=\phi_{n,\nu}(x)$ before and after the transition:
\begin{equation}
	\tau_{ij} = \int_0^\infty dx \; \phi_{i,1} (x) \phi_{j,0} (x).
\end{equation}
The $\tau_{ij}$ play an important role for strong vibrational coupling $\epsilon$. 
%
Here and in contrast to the usual small polaron models (linear oscillator coupling) %
\cite{koch,mitraPRB,Flensberg},
they have to be calculated numerically from the exact eigenfunctions of $H_{\rm dot}$. Some Franck-Condon factors are shown in Fig. \ref{figPotSpecFcDiam}(b). As a function of $\Omega$, there are two different regimes where either $\tau_{11}$ or $\tau_{21}$ is dominating. The sudden rise of $\tau_{11}$ is due to the level anti-crossing after which the shape of the first eigenstate resembles the harmonic oscillator ground state and vice versa for $\tau_{21}$ and the second eigenstate. Also the sign of the $\tau_{ij}$ has to be taken into account (cf. Eq. \ref{master1}) contrary to 
the usual rate equation approach 
where the Franck-Condon factors occur as squared values only.
Our numerical evaluation shows that the vibrational off-diagonal elements of Eq.(\ref{master1}) are not small in the regimes we are interested in and therefore must not be neglected.
 
\begin{figure}[t]\centering
		 \includegraphics[width=\columnwidth]{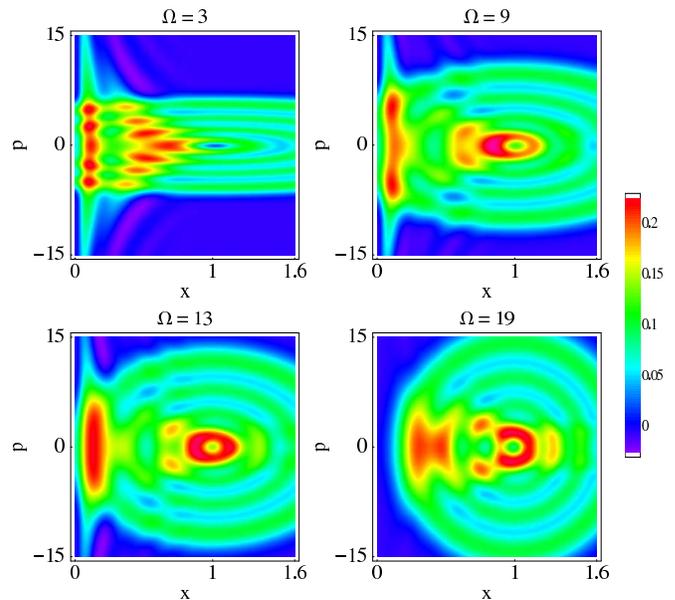}
		\caption{(color online). Wigner distribution of the dot-resonator for coupling strength $\epsilon=12$ and different parameters $\Omega$, cf. Eq. (\ref{parameters}). The contribution of the harmonic oscillator Wigner distribution (uncharged resonator states $|n,0\rangle$) is clearly visible as a series of concentric rings around the equilibrium point at $x=1$. For small $\Omega$,  the system is well localized close to the  surface, whereas for intermediate $\Omega$  it clearly lives in two separate regions of phase space corresponding to the two effectively localized two-level states. For larger $\Omega$, there is again localization around $x=1$. 
}\label{figWigner}
 \end{figure}

{\em Results .-}
Even without damping of the resonator (see below), our system has quite  a large parameter space ($\Omega, \epsilon, \mu_L, \mu_R, \Gamma_L, \Gamma_R$) and consequently exhibits a rich behaviour. We therefore first visualize the stationary state $\rho_{stat}=\rho(t\to \infty)$ of the dot-resonator system by plotting its Wigner transform, $W(x,p)= \frac{1}{2 \pi}\int_{-\infty}^\infty dy e^{ipy}\langle x+\frac{1}{2}y|\rho|x-\frac{1}{2}y\rangle$, which displays the expected behaviour as mentioned above:
the system clearly `lives' in two separate regions in phase space, one being directly near the surface ($x=0$) and well localized in $x$ direction, whereas the other region strongly depends on the parameters but is generally located around the equilibrium point $a$ (Fig. \ref{figWigner}). The concentric rings around this point are the contribution of the empty oscillator, which are just the usual harmonic oscillator Wigner functions. These contributions `interfere' with the contributions of the charged oscillator which, as a result, leads to the two-level characteristics visible in the total Wigner function. 

\begin{figure}[t]\centering
		\includegraphics[width=\columnwidth]{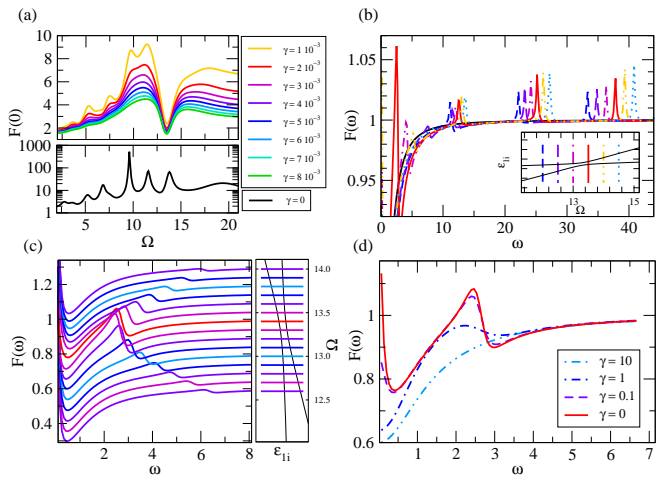}
		\caption{(color online). \textbf{(a)} Zero Frequency Fano factor for zero damping (lower frame) and several damping strength $\gamma \in [10^{-3},8\times10^{-3}]$ (upper frame). \textbf{(b)} Frequency dependent Fano factor $F(\omega)=S(\omega)/\langle \mathcal{I}\rangle$ for systems with eigenfrequencies $\Omega$ as indicated in the inset, $V_{Bias} = V_{Gate} = 30$. Coupling parameters are $\Gamma_L=1/2\pi$ and $\Gamma_R=1/4\pi$. For comparison the case where only diagonal elements of the Liovillian are considered (black line). \textbf{(c)} As (b). The sensitivity of the noise to the level anti-crossing is seen as a growing and flattening peak at the frequency corresponding to the level spacing for eigenfrequencies scanning through the anti-crossing point. The oscillator eigenfrequency of each line is indicated by the right frame as well as its position with respect to the avoided-crossing. The scale of the axis refers to the oscillator frequency at the level anti-crossing point $\Omega = 13.4$ (red line) the other lines have an offset of $\pm n 0.05$.  \textbf{(d)} $F(\omega)$ for $\Omega=13.4$ and several damping strength $\gamma$. For high damping the peak at the level spacing disappears and the noise spectrum resembles the single resonant level result.}\label{figNoise}
 \end{figure}

The stationary current, Fig. \ref{figPotSpecFcDiam}(c)\&(d) shows the usual signatures of a Franck-Condon blockade, i.e. a suppression of the low bias current that cannot be lifted by application of a gate voltage 
and which is due to the Franck-Condon factors $\tau_{ij}$ that describe the overlap of vibrational states before and after tunneling of electrons.  
Furthermore, we also observe negative differential conductance \cite{boeseEuPhys} (Fig. \ref{figPotSpecFcDiam}(d)), {and huge Fano factors} \cite{koch} (Fig. \ref{figNoise}(a)).

The most significant information about the resonator, however, can be 
directly 
extracted from the electronic noise spectrum $S(\omega)$. Following the method developed 
in 
\cite{novPhysicaE},
we calculate the frequency dependent noise from the convenient expression
\begin{equation}
S_{RR}(\omega) = \langle \mathcal{I}_R\rangle -2\rm{Re} \langle\mathcal{I}_R \mathcal{R}(\omega) \mathcal{I}_R\rangle,
\end{equation}
where $\mathcal{I}_R$ is the part of the Liouvillian that generates a quantum jump (an electron tunneling to the, e.g., right reservoir), $\mathcal{R}(\omega)$ is the pseudo-inverse of the Liouvillian $\mathcal{L}$ and $\langle \bullet \rangle = Tr[\bullet \rho_{stat}]$.
Note that for simplicity, we only calculate the noise contribution in the right reservoir which (together with the contribution from the left reservoir) is sufficient to obtain the noise for strongly asymmetric capacitance coefficients.

 Figs. \ref{figNoise}(b)\&(c) demonstrate that the noise is indeed extremely sensitive \cite{aguadoPRL} to changes in the energy level spectrum of the resonator. The level splitting close to the avoided level crossing becomes visible as the appearance of a feature in $S(\omega)$ at frequencies $\omega$ close to the corresponding Bohr frequencies (level splitting in the vibrational spectrum, Fig. \ref{figNoise}(b) inset). The electronic noise thus acts as a direct probe of vibrational coherence. We point out that the appearance of non-trivial spectral information in $S(\omega)$ is due to retaining the non-diagonal elements in the Master equation Eq. (\ref{master1}). In order to emphasize this point, we have compared our results with a corresponding calculation where these elements were set to zero (Fig. \ref{figNoise}(b)).

In order to elucidate the role of damping, we again sandwiched the usual Lindblad form $\mathcal{L}_{\rm damp}=\frac{\gamma}{2} (2 a\rho a^{\dagger} - \rho a^\dagger a - a^\dagger a \rho)$ with the exact vibrational eigenfunctions. Here, $\gamma$ is the damping rate of the uncharged oscillator, described by $a$ and $a^\dagger$. The most dramatic effect this damping has on the system is a considerable reduction of the zero frequency Fano factor (Fig. \ref{figNoise} (a)). 
The huge zero frequency Fano factor  can be understood as a dynamical channel blockade as described in \cite{koch}: The peaks observed in Fig. \ref{figNoise}(a) all occur at $\Omega$ values where at least one Franck-Condon factor is zero (cf. Fig. \ref{figPotSpecFcDiam}(b)). This means one transport channel is blocked and the current is withheld for some time until it can flow through some other channel. The peak at $\Omega=9.6$ is unusually high compared with the other because here several Franck-Condon factors are zero or of the order of $10^{-2}$. The damping now has the effect of allowing relaxations between the levels and thus even for very small damping the channel blockade is lifted.

The observed resonance peak at the level spacing energies is also suppressed for a system subjected to damping (Fig. \ref{figNoise}(a)), 
and the quantum coherence between the vibrational states is gradually lost.
For high damping constants $\gamma$ the whole noise spectrum approaches the limit of a static quantum dot with two levels 
\cite{belzig}, as one would expect. 

Finally, we comment on possible experimental realizations of our predictions. 
The two model parameters $\Omega$ and $\epsilon$,  Eq. (\ref{parameters}), fix the ratio $M/\omega_0=\hbar^3(8\pi\varepsilon_0\varepsilon_r/e^2)^2\epsilon^2/\Omega$ between effective oscillator mass $M$ and frequency $\omega_0$. 
The values $\Omega=13.4$ and $\epsilon=12$ used above yield a distance $a\sim 10^{-9}$m and would in fact correspond to an `electron droplet' oscillator of a  few electron masses, i.e. quantum dot electrons collectively oscillating at frequencies $\omega_0\sim 10^{14}$Hz (using $\epsilon_r \sim 10$). In calculations for larger  $\epsilon$, however, we also confirmed  that one can easily reach $M$ of the order of a few proton masses or larger, with the two-level approximation still being valid. These masses are reasonable for vibrating molecular junctions or even larger macroscopic oscillators. Since $\Omega$ is proportional to the distance $a$ it could be tuned by using a metallic STM tip as the part of the environment were the image charge occurs. Then one would be able to scan $\Omega$ and thus find the two level parameter regime, where the described noise features can be observed.

We acknowledge useful discussions with F. Haupt. This work was supported by DFG project BR/1528/5-1 and the WE-Heraeus foundation.


\end{document}